\documentclass[a4paper]{article}

\usepackage[utf8]{inputenc}
\usepackage[T1]{fontenc}
\usepackage{graphicx}
\usepackage{hyperref}
\usepackage{amsmath}
\usepackage{amssymb}
\usepackage{mathenv}
\usepackage{multirow}

\DeclareMathAlphabet{\itbf}{OML}{cmm}{b}{it}

\numberwithin{equation}{section}

\newcommand{\R}{\mathbb{R}}

\newcommand{\I}{\mathcal{I}}
\newcommand{\E}{\itbf{E}}
\newcommand{\B}{\itbf{B}}
\newcommand{\jj}{\itbf{j}}
\newcommand{\bx}{\itbf{x}}
\newcommand{\bz}{\itbf{z}}

\newcommand{\pOmega}{\partial \Omega}
\newcommand{\pD}{\partial D}
\newcommand{\supp}{\textrm{supp}}
\newcommand{\norm}[2]{\| #1 \|_{#2}}
\newcommand{\abs}[1]{\vert #1 \vert}
\newcommand{\ie}{\emph{i.e.}~}
\newcommand{\Div}{\nabla \cdot}

\newcommand{\ddn}[1]{\frac{\partial #1}{\partial \nu}}
\newcommand{\Spm}{\textrm{S} \cdot \textrm{m}^{-1}}
\newcommand{\pd}{\itbf{p}_D}
\newcommand{\komega}{k(\omega)}
\newcommand{\kDomega}{k_D(\omega)}
\newcommand{\ktildeomega}{\tilde{k}(\omega)}
\newcommand{\MkD}{{\bf M}(\kDomega,D)}

\newcommand{\bnu}{{\boldsymbol{\nu}}}

\title{Mathematical modelling of the electric sense of fish: the role of multi-frequency measurements and movement}
\author{Habib Ammari, Thomas Boulier, Josselin Garnier, Han Wang}
\date{}

\begin{document}

\maketitle

\begin{abstract}
Understanding active electrolocation in weakly electric fish remains a challenging issue. In this article we propose a mathematical formulation of this problem, in terms of partial differential equations. This allows us to detail two algorithms: one for localizing a target using the multi-frequency aspect of the signal, and antoher one for identifying the shape of this target. Shape recognition is designed in a machine learning point of view, and take advantage of both the multi-frequency setup and the movement of the fish around its prey. Numerical simulations are shown for the computation of the electric field emitted and sensed by the fish; they are then used as an input for the two algorithms.
\end{abstract}

\section{Introduction}
\label{sec:introduction}

In order to build a proper artificial electric sense, it is essential to understand how it works in biology. In this regard, the study of active electrolocation in weakly electric fish is crucial. Indeed, from an electric potential of about $1 \, \textrm{mV}$ oscillating at about $1$--$10 \, \textrm{kHz}$, weakly electric fish are able to recognize an object in their surrounding (for a review, see~\cite{moller1995electric} and references therein). Hence, they are undoubtly a source of great inspiration for neuro-ethology, underwater robotics, signal processing, as well as applied mathematics.

Several species of fish share this remarkable sense. They are classified in various families, all belonging to two different orders: Gymnotiforms in South America, and Mormyriforms in Africa. Moreover, according to the time representation of their Electric Organ Discharges (EODs), they are also divided into two types: wave-type species (such as \emph{Apteronotus albifrons}) and pulse-type species (\emph{e.g.} \emph{Gnathonemus petersii}). Known for several centuries, the electrogenesis and electroreception abilities of the wider set of species called \emph{electric fish} have been studied extensively~\cite{finger2011shocking}. In 1958, Lissmann and Machin showed that for the weakly electric fish, this ability is used for electrolocation instead of hurting preys~\cite{lissmann1958mechanism}. Furthermore, they gave the physical principles on which relies this electric sense: using cylinders in the water tanks of their \emph{Gymnarchus niloticus}, they showed that these objects disturb the self-emitted electric field like an electric dipole. The electroreceptors allow the fish to distinguish such a difference, which in turn is a clue for electrolocation. However, one important question remains: how to estimate the location of the object from the measurement of this difference?

Experimental, modelling and numerical approaches have been carried since this discovery by Lissman and Machin. From behavioral studies, we now know that these fish are able to estimate the distance~\cite{von1998electric}, recognize the shape~\cite{von2007distance}, and the electric capacitance and conductivity~\cite{von1999active} of an object in their surrounding. In these studies, a fish is placed in front of two doors, each one hiding a different object. The fish is then trained to chose one of the two objects, in a reward/punish setup. More theoretically, the electric dipole formula has been investigated in more details. Indeed, Bacher in 1983~\cite{bacher1983new} argued that the electric dipole formula given by Lissmann and Machin did not explain the phase shift observed when the electric premittivity of the object differs from the electric permittivity of the water, although this phase difference is measured acurately by some species such as in the \emph{Eigenmannia} genus~\cite{rose1985temporal}. Then, in 1996 Rasnow~\cite{rasnow1996effects} solved this issue by considering a complex-valued conductivity; he also extended the range of shapes on which the dipolar approximation can be applied. From the numerical simulation point of view, various works have been carried since the 70's, for example finite differences schemes in 1975 by Heiligenberg~\cite{heiligenberg1975theoretical} and finite elements in 1980 by Hoshimiya \emph{et al.}~\cite{hoshimiya1980theapteronotus}. In the latter, a simplified geometry of fish was used: it was represented as an ellipse, divided into two areas (the low conductive and thin skin, and the body). Their aim was to optimize the non-uniform values of the skin's conductivity, and by optimizing it according to experimentally measured field, they conclude that the tail region is more conductive than the head region. These models were then improved, as can be seen in~\cite{babineau2006modeling,maciver2001computational,migliaro2005theoretical}. Finally, in the 90's Chris Assad considered a boundary element method to solve this numerical simulation issue~\cite{assad1997electric}. His model took into account the highly resistive and thin skin, as well as the time dependance of the EOD. He compared his model to several \emph{in vivo} experiments involving different species of fish~\cite{rasnow1988simulation}.

Those advances in the field of neuro-ethology inspired researchers in robotics to develop underwater probes and sensors in order to effectively navigate with the help of this electric sense. Mainly two teams gather the most research about this stimulating subject: one in Nantes lead by Frédéric Boyer~\cite{boyer2012model}, and another in Chicago lead by Malcolm MacIver~\cite{maciver2004designing}. Just to mention a few of their studies, they face important challenges such as target location~\cite{lebastard2013environment,solberg2008active}, shape recognition~\cite{bai2015finding,lanneau2016object}, autonomous (or reactive) navigation~\cite{boyer2013underwater}, or docking~\cite{boyer2015underwater}. We leave the interested reader to these articles and the references therein for a more complete review of this area. Let us mention however that for these works, there is a need for more quantitative assessments of the electric sense: how precisely an object disturb the electric field, and how to compute back its location and shape.

Mathematically speaking, this is called an \emph{inverse problem}, as opposed to a \emph{forward problem}. The latter would be to compute the electric field surrounding the fish, knowing everything about the object (position, shape, material). On the contrary, the inverse problem here is to recover as much information as possible about the object from the knowledge of the electric field at the surface of the fish's skin. Given the low frequencies of emission (see Section~\ref{sub:quasi_static_approximation}), this problem lies in the domain of Electrical Impedance Tomography (EIT). EIT is a non-invasive imaging technique in which an image of the conductivity or permittivity of a medium is inferred from surface electrode measurements. It is studied as a non-invasive imaging technique, in particular for medical imaging, non-destructive testing, or geophysical probing (see reviews in~\cite{borcea02,cheney1999electrical}). In the mathematical literature it is also known as Calder\'on's problem from Calder\'on's pioneer contribution~\cite{calderon80}. This type of problem is \emph{ill-posed}, in the sense that existence, uniqueness, or continuity of the solution  is not guaranteed~\cite{uhlmann09}. The resolution of the inverse problem is often formulated as a minimization problem, which consists in minimizing the error between the measured data and the synthetic data obtained by solving numerically the forward problem with a candidate object. It requires careful discretization and regularization and sometimes numerically intensive calculations, however, it is known to be sensitive to noise and to have poor spatial resolution~\cite{borcea02,brown03}.

Since 2010, our team has been working on the mathematical modelling of active electrolocation. Indeed, having heard about that the electric fish are able to solve --~in some way~-- the Calder\'on's problem raised our curiousity. In other words, the problem of recovering the object from what the fish can feel is a very difficult problem, thus making it even more fascinating as the fish seem to perform better than the most recent medical EIT devices which typically consist of a few tens of electrodes only. Hence, we wanted to have equations governing the electric field surrounding the fish (\emph{i.e.} a model of the forward problem), so that we could imagine original solutions of the inverse problem. Our inspiration largely took its source in the aformentioned works, and relevent references will be done throughout the text. The aim of this article is to summarize our works on target location estimation~\cite{ammari2013modeling} and shape identification~\cite{ammari2014shape}, as well as making connections between our theoretical studies and what it is intended to model. We will show that it is possible to extract some information about the object in a quite robust and straightforward manner, even with less than one hundred electrodes. We wanted to make accessible some mathematical concepts that could be useful to researchers in biology and robotics, in order to engage further discussion and collaboration.

The outline of this paper is as follows. In Section~\ref{sec:forward_problem}, we derive the equations governing the electric field emitted by the fish and show numerical simulations. In Section~\ref{sec:inverse_problem}, we explain the localization (Section~\ref{sub:localization}) and shape recognition (Section~\ref{sub:shape_recognition}) algorithms, which both rely on a formula called dipolar approximation (Section~\ref{sub:dipolar_approximation}).

\section{Forward Problem}
\begin{figure}[!ht]
	\centering
	\includegraphics[width=10cm]{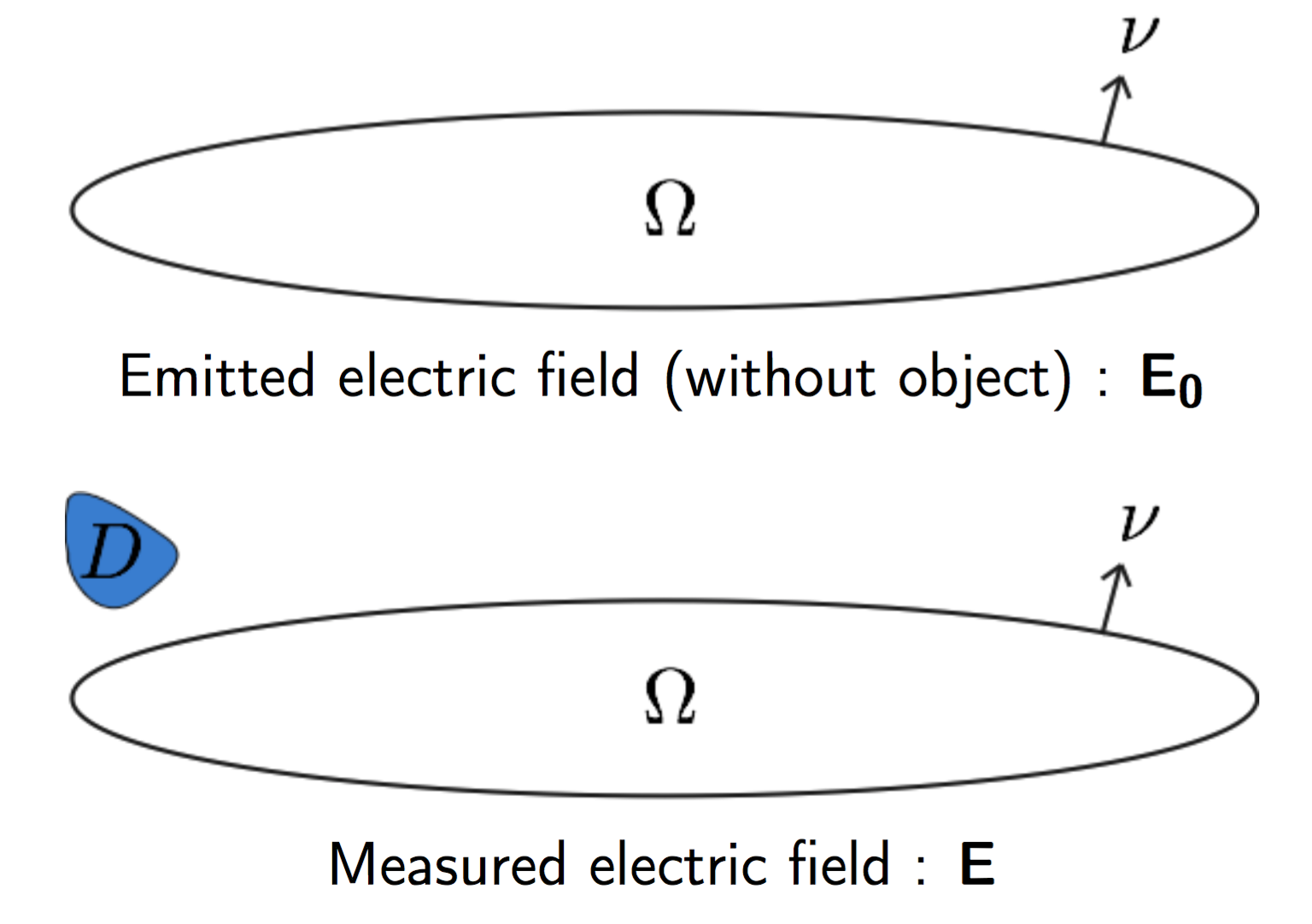}
	\caption{Problem considered here: knowing the electric current $(\E-\E_0) \cdot \bnu$ at the surface of the skin $\pOmega$, how can we determine the target $D$?\label{fig:model}}
\end{figure}

{}
\label{sec:forward_problem}
Let us consider the model depicted in Figure~\ref{fig:model}. The body of the fish is modelled as a domain $\Omega \subset \R^d$, where $d \in \{2,3\}$. For the sake of clarity, we plot the results for $d=2$ in this paper.
This mathematical model is of course extremely simplified compared to the complexity of a fish revealed by biology,
but our objective is to extract the few ingredients that are sufficient to explain the fish electric sense, as was done for example in the numerical simualtions of Hoshimiya \emph{et al.}~\cite{hoshimiya1980theapteronotus}.

The goal is to recover the object $D$, another set of $\R^d$ which is away from $\Omega$.
The localization is explained in Section~\ref{sub:localization} whereas the shape identification is shown in Section~\ref{sub:shape_recognition}.

In this section, we focus on the so-called \emph{forward problem}: what are the equations governing the transdermal electric current $(\E-\E_0) \cdot \bnu$ on $\pOmega$ (Section~\ref{sub:quasi_static_approximation} and Section~\ref{sub:boundary_conditions}), and how can we compute it numerically (Section~\ref{sub:numerical_simulations})? The results presented here can be found with more mathematical details in~\cite{ammari2013modeling}.

\subsection{Quasi-Static Approximation}
\label{sub:quasi_static_approximation}
In this subsection we show that, given the low frequency of the electric field, it can be derived as a complex-valued electric potential.

From this point, let us precise that we will work in the frequency domain, so that time derivatives will be simplified by Fourier transform. It has an impact of what we are modelling exactly, \emph{i.e.} from now on we can only discuss about wave-type species such as \emph{Apteronotus albifrons}. For example, after discretization, we would have a discrete set of frequencies (the fundamental and its harmonics), see for example~\cite[fig. 12.3]{evans2006physiology} for such a transform in \emph{Eigenmannia virescens}. These frequencies are known to remain centred on a specific frequency unless their dominance status changes, or the temperature or pH of the water varies~\cite{dunlap1998diversity}.

Having this in mind, let us start with the Maxwell system:
\begin{align}
  	\Div \E  &= \frac{\rho}{\varepsilon}, \label{eq:Maxwell1} \\ 
  	\Div \B  &= 0, \label{eq:Maxwell2} \\ 
  	\nabla \times \E &= - i \omega \B, \label{eq:Maxwell3} \\ 
  	\nabla \times \B &= \mu (\jj_i + \jj_s + i\omega \varepsilon \E), \label{eq:Maxwell4}
\end{align}
where $\varepsilon$ (resp. $\mu$) is the electric permittivity (resp. the magnetic susceptibility) of the medium
and $\omega$ is the frequency. 
The sources are $\rho$ (density of electric charges) and $\jj_s$ (density of electric current). 
Whereas the former are null, the latter need to be considered carefully. 
Indeed, the current density comes from the electric organ of the fish. 
It is usually a long filament at the posterior part of the body~\cite{moller1995electric}. 
In any case, it can be modelled as a distribution contained into the body:  $\supp (\jj_s) \subset \Omega$. 
Finally, the electrical current $\jj_i$ in (\ref{eq:Maxwell4}) is the one induced by Ohm's law,
\begin{equation}
	\jj_i = \sigma \E,
\end{equation}
where $\sigma$ is the conductivity of the medium.

The electro-quasistatic~(EQS) approximation states that, if the wavelength $\lambda$ is large compared to the typical length of the problem $L$, then $\E$ can be considered as irrotationnal, \ie the right-hand side of (\ref{eq:Maxwell3}) is neglected~\cite{klinkenbusch2011domains}. In other words, if $\lambda \gg L$, then $\nabla \times \E \approx 0$. In nature emission frequencies are always below $10$kHz \cite{nelson2006sensory},
which means that $\lambda$ is always larger than $10\textrm{km}$, which is much larger than $L$ if we consider it as the typical size of the fish; indeeed, the electrolocation range is known to be one body-length at maximum~\cite{moller1995electric}. Hence, the EQS approximation is very well suited in our case and thus we can write $\nabla \times \E = 0$ instead of (\ref{eq:Maxwell3}). Then, it follows that there exists a frequency-dependent, complex-valued potential $u$ such that
\begin{equation}
	\E = \nabla u.
\end{equation}
Taking the divergence of~(\ref{eq:Maxwell4}), we finally obtain
\begin{equation}
	\Div \big( \komega \nabla u \big) = f, \label{eq:conductivity}
\end{equation}
where we have defined $f = -\Div \jj_s$ as the source coming from the electric organ, and $\komega = \sigma + i \varepsilon \omega$ as the complex-valued conductivity; this complex-valued conductivity is the same idea that was used by Rasnow in~\cite{rasnow1996effects}.

To conclude, the equation~(\ref{eq:conductivity}) is the one governing the transdermal electric current. If we note $U$ the potential associated to the background electric field ($\E_0 = \nabla U$), the transdermal electric current stated in Figure~\ref{fig:model} is translated into
\begin{equation}
	(\E-\E_0) \cdot \bnu = \ddn{u} - \ddn{U}.
\end{equation}

\subsection{Boundary Conditions}
\label{sub:boundary_conditions}
\begin{figure}[!ht]
	\centering
	\includegraphics[width=10cm]{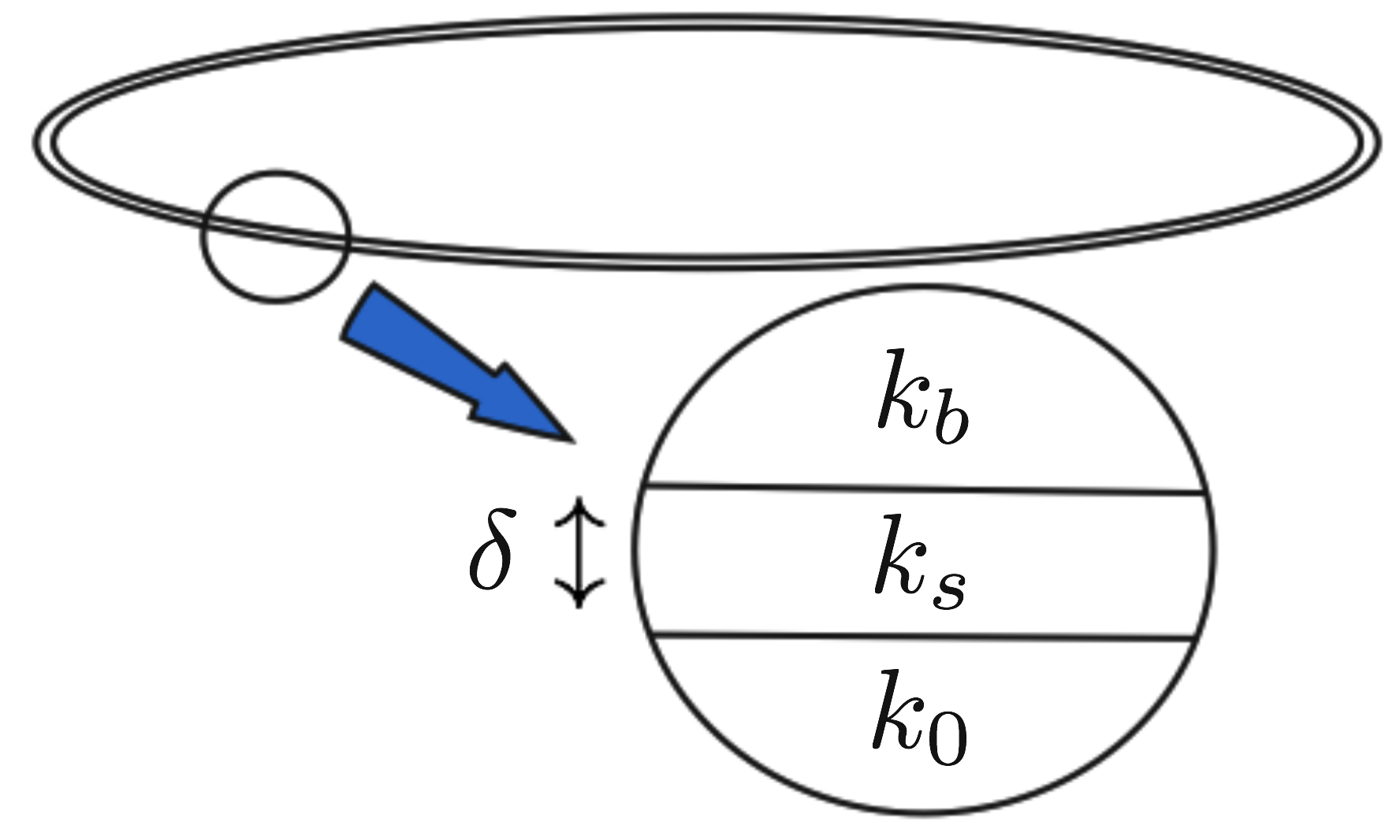}
	\caption{Boundary conditions over the skin.}
	\label{fig:skin}
\end{figure}

Now that we have the partial differential equations governing the electric field, given in~(\ref{eq:conductivity}), let us focus on the boundary conditions.
They strongly depend on the distribution of the conductivity, $\komega$, which can be modelled as piecewise constant: $k_w$ in the water, $k_D$ in the object, $k_b$ in the body of the fish, and $k_s$ in the skin. Note that since the water, the body, and the skin are not dielectric materials, we have
\begin{equation*}
	\Im(k_w) = \Im(k_b) = \Im(k_s) = 0.
\end{equation*}
Hence, two boundary conditions appear to matter: at the surface of the object $\pD$ and over the skin $\pOmega$. \\
1) The boundary conditions at the surface of the object can be addressed easily. Indeed,
the piecewise constant conductivity imposes the following jump relations for $\bx \in \pD$~\cite{ammari2004reconstruction}
\begin{align}
	u(\bx^-) &= u(\bx^+), \label{eq:BC_D_1} \\
	k_w \ddn{u}(\bx^-) &= k_D(\omega)\ddn{u}(\bx^-). \label{eq:BC_D_2}
\end{align}
The notation $\bx^{\pm}$ means the inner/outer limit at the boundary of $\pD$. 
More precisely, for a function $w$ defined on $\R^d$, one has
\begin{equation}
	w(\bx^{\pm}) = \lim_{h \rightarrow 0} w(\bx \pm h \bnu), \; \; \bx \in \pD,
\end{equation}
where $\bnu$ is the outward normal unit vector of $\pD$.\\
2) The boundary conditions over the skin are a bit more complicated (see Figure~\ref{fig:skin}).
This is due to the fact that, compared to the water which has a conductivity of the order of~$0.01 \Spm$~\cite{maciver2001prey}, the skin is very resistive ($10^{-4} \Spm$~\cite{budelli2000electric}) and the body is very conductive ($1 \Spm$)~\cite{scheich1973coding}. In other words, one has
\begin{equation}
	k_s \ll k_w \ll k_b.
\end{equation}
Futhermore, the skin is very thin: if we denote its thickness by $\delta$, we have~\cite{zakon1986electroreceptive}
$$ \delta \approx 100 \mu\textrm{m} \ll L,$$
where $L$ was defined as the body length in Section~\ref{sub:quasi_static_approximation}.
In~\cite{ammari2013modeling} we have shown in the case $d=2$ that, when $\delta/L \ll 1$ and $k_s/k_w \ll 1$ ,
but $\delta k_w / (L k_s)$  is of order one (or smaller),
we have the following effective relation for $\bx \in \pOmega$:
\begin{equation}
	u(\bx^+) - u(\bx^-) = \xi \ddn{u}(\bx^+), \label{eq:robin_BC}
\end{equation}
where $\xi = \delta k_w/k_s$ is called the \emph{effective thickness} in Assad's work~\cite{assad1997electric}. Indeed, equation~(\ref{eq:robin_BC}) is exactly the same as the one used in his model.
On the other side the limit $k_b / k_w \gg 1$ gives
\begin{equation}
	\ddn{u}(\bx^-) = 0. \label{eq:neumann_BC}
\end{equation}
To get a well-posed problem, we should add the far field condition $u(\bx) = O(|\bx|^{1-d})$ as $|\bx|\to \infty$, if the problem is formulated in an open medium, or any prescribed condition corresponding to the experimental configuration.

\subsection{Numerical Simulations}
\label{sub:numerical_simulations}
Taken altogether, we have to solve a system composed by the partial differential equation (\ref{eq:conductivity}) with boundary conditions (\ref{eq:BC_D_1}), (\ref{eq:BC_D_2}), (\ref{eq:robin_BC}), and (\ref{eq:neumann_BC}). Hence, $u$ is solution of the following system
\begin{align}
	\Div \big( \komega \nabla u \big) &= f, \label{eq:u_first} \\ 
	u(\bx^-) &= u(\bx^+), & \bx \in \pD \\
	k_w \ddn{u}(\bx^-) &= k_D(\omega)\ddn{u}(\bx^-), & \bx \in \pD \\
	u(\bx^+) - u(\bx^-) &= \xi \ddn{u}(\bx^+), & \bx \in \pOmega\\
	\ddn{u}(\bx^-) &= 0, & \bx \in \pOmega \label{eq:u_last}
\end{align}
and the background solution $U$ is given by
\begin{align}
	\Div \big( \ktildeomega \nabla U \big) &= f, \label{eq:U_first}\\
	U(\bx^+) - U(\bx^-) &= \xi \ddn{U}(\bx^+), & \bx \in \pOmega\\
	\ddn{U}(\bx^-) &= 0, & \bx \in \pOmega \label{eq:U_last}
\end{align}
where $\ktildeomega$ is equal to $k_b$ in the body of the fish $\Omega$, and $k_w$ outside, \emph{i.e.} in the water.

\begin{figure}[!ht]
	\centering
	\includegraphics[width=10cm]{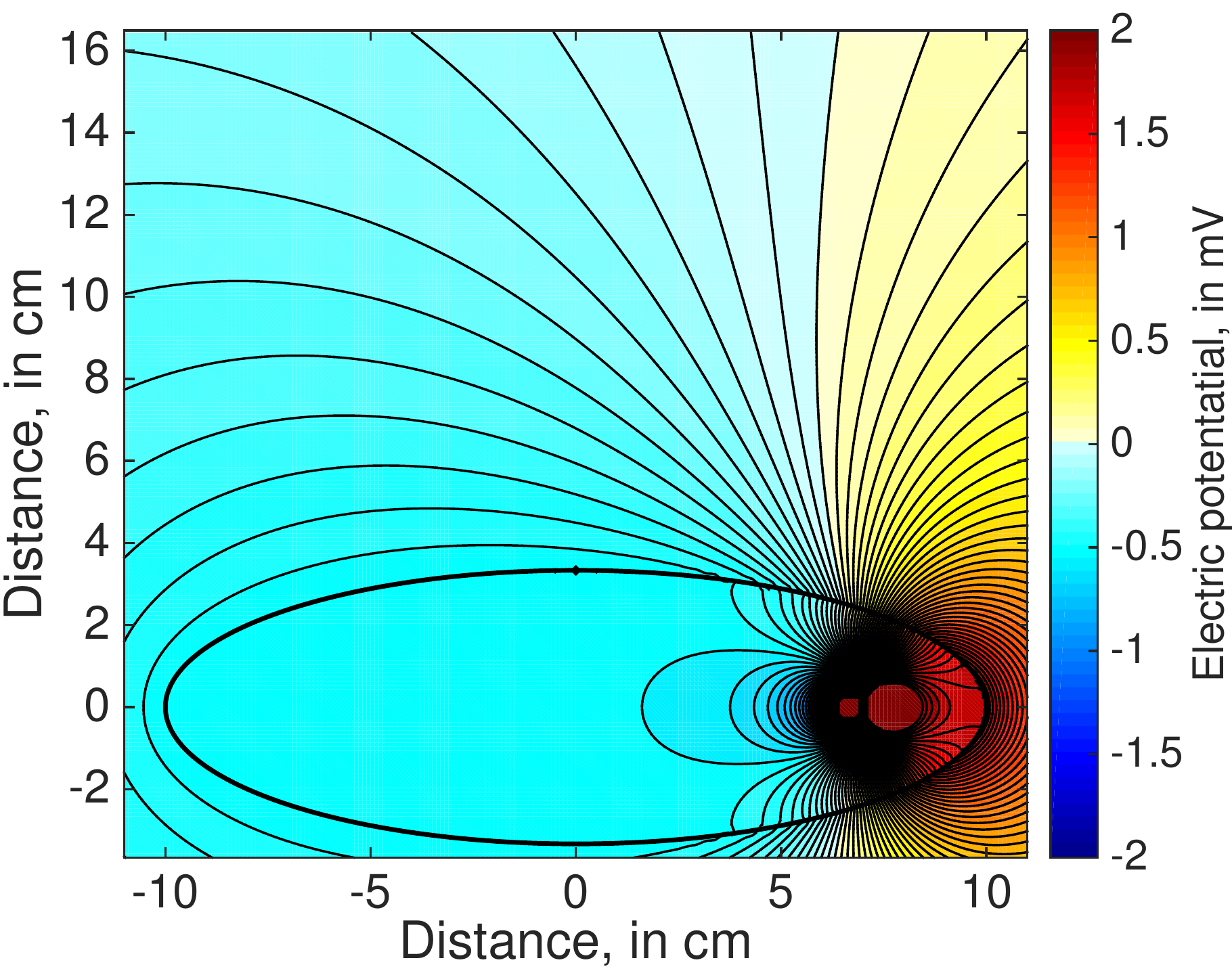} \\
	\includegraphics[width=10cm]{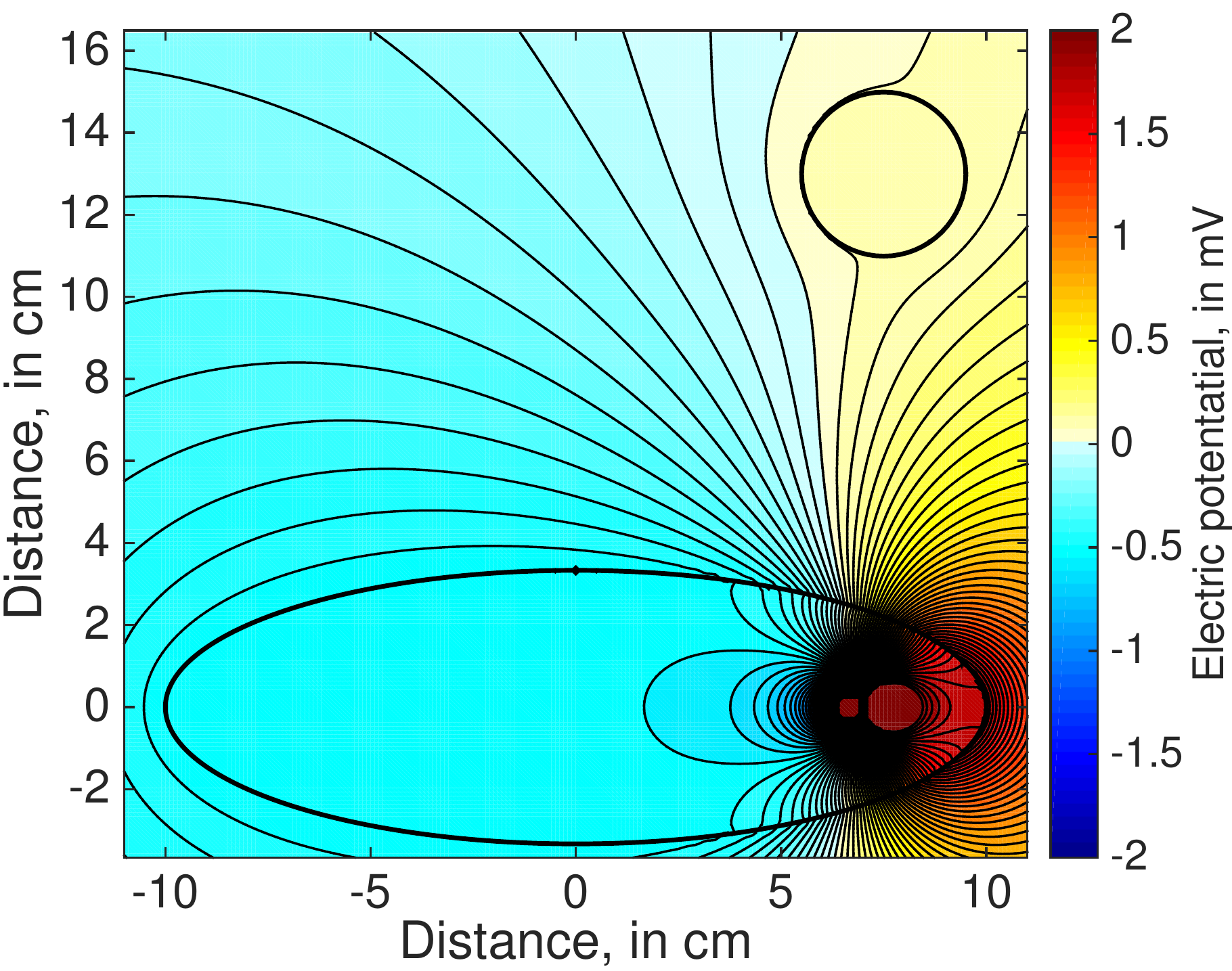}
	\caption{Numerical simulations for the background electric potential $U$ -~system (\ref{eq:U_first}-\ref{eq:U_last}), upper image~- and for the perturbed electric potential $u$ -~system (\ref{eq:u_first}-\ref{eq:u_last}), lower image.}
	\label{fig:numerics}
\end{figure}

Using layer potential representations for the solutions of the systems (\ref{eq:u_first}-\ref{eq:u_last}) and (\ref{eq:U_first}-\ref{eq:U_last}), we have developed a MATLAB script for their numerical approximations\footnote{This script is now part of the package SIES (\emph{Shape Identification in Electro-Sensing}), which can be found at \url{https://github.com/ens2013/SIES/}}, using boundary element methods like in Assad's thesis~\cite{assad1997electric}. In Figure~\ref{fig:numerics}, we have plotted the solutions $u$ and $U$ when $d=2$, the body $\Omega$ is an ellipse, the object $D$ is a disk, and the source $f$ is a dipole.

\section{Inverse Problem}
\label{sec:inverse_problem}
In the previous section, we have derived the equations governing the transdermal electric current at the surface of the skin, and we have shown results of numerical simulations.
In this section, we show the two main results of our studies: how to localize the object $D$ based on the knowledge of $\ddn{u}-\ddn{U}$ (Section~\ref{sub:localization}), and how to recognize its shape when the fish has already memorized several objects (Section~\ref{sub:shape_recognition}). These methods are based on a dipolar approximation of the solution $u$, which is explained in Section~\ref{sub:dipolar_approximation}. 

\subsection{Dipolar Approximation}
\label{sub:dipolar_approximation}
The dipolar approximation states that if $D$ is small enough, the difference $u-U$ can be expressed as the electric potential coming from an electric dipole centered in $D$. In this subsection we only present this result numerically, in order to make things more intuitive. For a complete proof of the formula in this context, see~\cite{ammari2013modeling}; for a more detailed review of the dipolar approximation in general, see the book~\cite{ammari2007polarization}.

\begin{figure}[!ht]
	\centering
	\includegraphics[width=10cm]{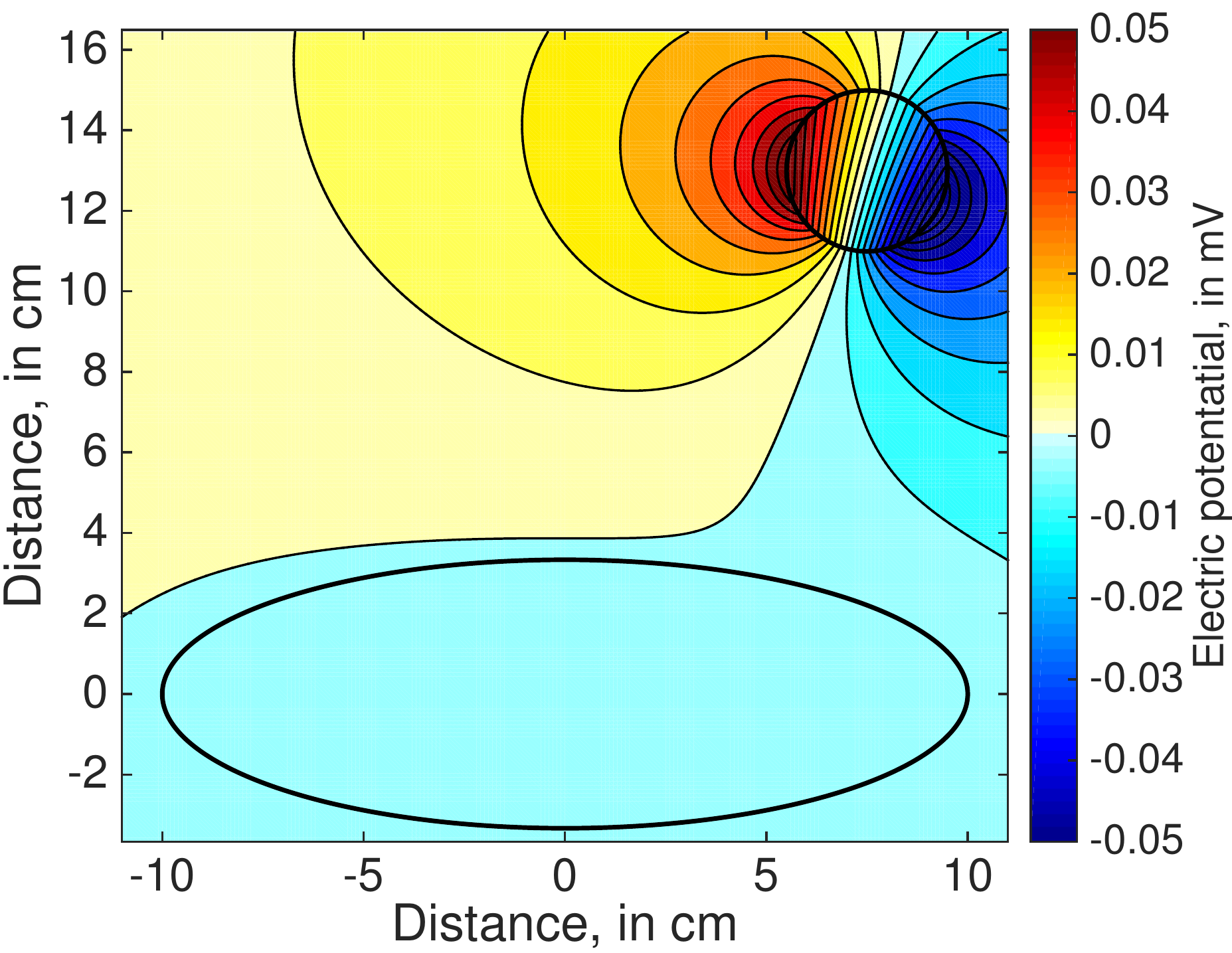}
	\caption{The difference between $u$ and $U$, taken from Figure~\ref{fig:numerics}.}
	\label{fig:u-U}
\end{figure}

Let us get back to the example in Section~\ref{sub:numerical_simulations}. In Figure~\ref{fig:u-U}, we have plotted the difference $u-U$. Qualitatively, we can see that this electric potential looks like the one emitted by an electric dipole coming from $D$.

Quantitatively, we have shown in~\cite{ammari2013modeling} that, if $D$ is small enough and sufficiently away from $\Omega$, then the fish feels a distorsion that is similar to the one produced by a dipole $\pd$. More precisely, we have
\begin{equation}
	\ddn{u}(\bx) - \ddn{U}(\bx) \approx \pd \cdot \nabla G(\bx-\bz), \quad \bx \in \pOmega, \label{eq:dipolar_formula}
\end{equation}
where $\bz$ is the center of mass of $D$ and $G$ is  the Green function for the Laplacian in
$\R^d$ (for instance $G(\bx) = \log (|\bx|)/(2\pi)$ in dimension $d=2$
or $G(\bx) = -1/(4\pi |\bx|)$ in dimension $d=3$). The vector $\pd$ is called the equivalent dipole, and it is given by
\begin{equation}
	\pd \approx \MkD \nabla U(\bz), \label{eq:equivalent_dipole}
\end{equation}
where $\MkD$ is a $d \times d$ complex-valued matrix that depends only on the shape $D$ of the object and its complex conductivity $\kDomega$, called the \emph{first-order polarization tensor}~\cite{ammari2007polarization}. This matrix maps the illuminating electric field $\nabla U(\bz)$ to the equivalent dipole $\pd$. When the conductivity $\kDomega$ is real, in other words when $\Im(\kDomega)=0$ and thus $\omega$ does not play any role anymore, one can find an ellipse or ellipsoid $\mathcal{E}$ such that
\begin{equation}
	\MkD = {\bf M}(k,\mathcal{E}).
\end{equation}
This means that the equivalent dipole would always be the same as the equivalent dipole of this ellipse 
or ellipsoid $\mathcal{E}$. This latter is then called the \emph{equivalent ellipse}~\cite{ammari2007polarization}.
Note, however, that when $\kDomega$ is frequency-dependent, the information that can be extracted is much richer.
For single-frequency data, it is only possible to identify a few characteristics of the object.  
When multi-frequency data are available, it is possible to get a lot of information about the object from the frequency
dependence of the observed first-order polarization tensors.
The need for multi-frequency data that we exhibit is in agreement with the complex emission patterns (pulse and wave) by fish.

\subsection{Localization}
\label{sub:localization}
\begin{figure}[!ht]
	\centering
	\includegraphics[width=10cm]{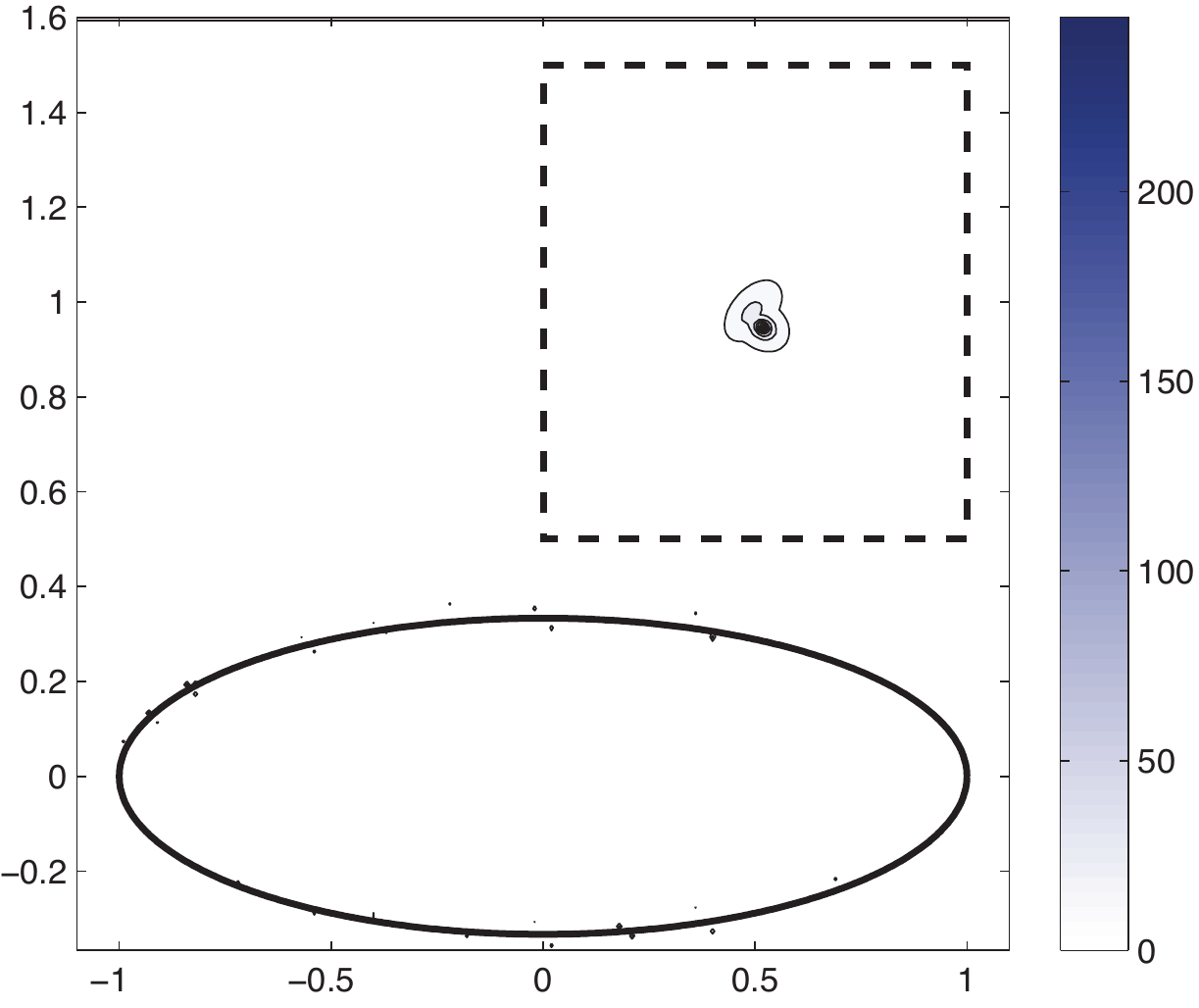}
	\caption{From the example shown in Figures~\ref{fig:numerics}-\ref{fig:u-U}, plot of the imaging functional presented in~\cite{ammari2013modeling}.}
	\label{fig:SF-MUSIC}
\end{figure}

In this subsection, we show that the multi-frequency aspect of the measurements are sufficient to localize an object with precision. For the sake of simplicity we focus our attention on the example that gave Figures~\ref{fig:numerics}-\ref{fig:u-U} in dimension $d=2$. Indeed, the particular case of $D$ being a disk is easier since we have~\cite{ammari2007polarization}
\begin{equation}
	\MkD = \abs{D} \frac{\kDomega-1}{2\kDomega+1} {\bf I}_2,
\end{equation}
where $\abs{D}$ denotes the volume of $D$ and ${\bf I}_2$ is the identity matrix.\footnote{In $\R^3$, we have a similar formula, \ie $\MkD$ is proportional to identity~\cite[p. 83]{ammari2007polarization}.}
Hence, equation~(\ref{eq:dipolar_formula}) becomes
\begin{equation}
	\ddn{u}(\bx,\omega) - \ddn{U}(\bx,\omega) = \abs{D} \frac{\kDomega-1}{2\kDomega+1} \nabla U(\bz) \cdot \nabla G(\bx-\bz),
	\label{eq:dipolar_formula_fish}
\end{equation}
which is equivalent to Lissman-Machin~\cite{lissmann1958mechanism} or Rasnow~\cite{rasnow1996effects} formulas, showing that~(\ref{eq:dipolar_formula}) is a generalization of these latters.
Hence, from~(\ref{eq:dipolar_formula_fish}) we can easily extract $\nabla U(\bz) \cdot \nabla G(\bx-\bz)$. The reason why we can recover the location $\bz$ of $D$ is that the function $\bz \mapsto \nabla U(\bz) \cdot \nabla G(\bx-\bz)$ is one-to-one. Therefore, it is possible to build an \emph{imaging functional} from the measured data, \ie a function $\bz_s \mapsto \I(\bz_s)$ that has a strong peak at $\bz_s = \bz$ (see Figure~\ref{fig:SF-MUSIC}).

\subsection{Shape Recognition}
\label{sub:shape_recognition}
Once the object $D$ is located (\ie we know $\bz$), we would like to extract $\MkD$ from the measurements. Indeed, we know that this polarization tensor contains all the necessary information about the shape of $D$~\cite{ammari2016shape}. More precisely, the function $\omega \mapsto \MkD$ uniquely determines $D$ in some class of domains.
However, it is not straighfoward to determine $D$ from $\MkD$.
In other words, the shape recognition problem has been reduced to a simpler, however still difficult, inverse problem: determine the shape of the object from the observed first-order polarization tensors.

\begin{figure}[!ht]
	\centering
	\includegraphics[width=10cm]{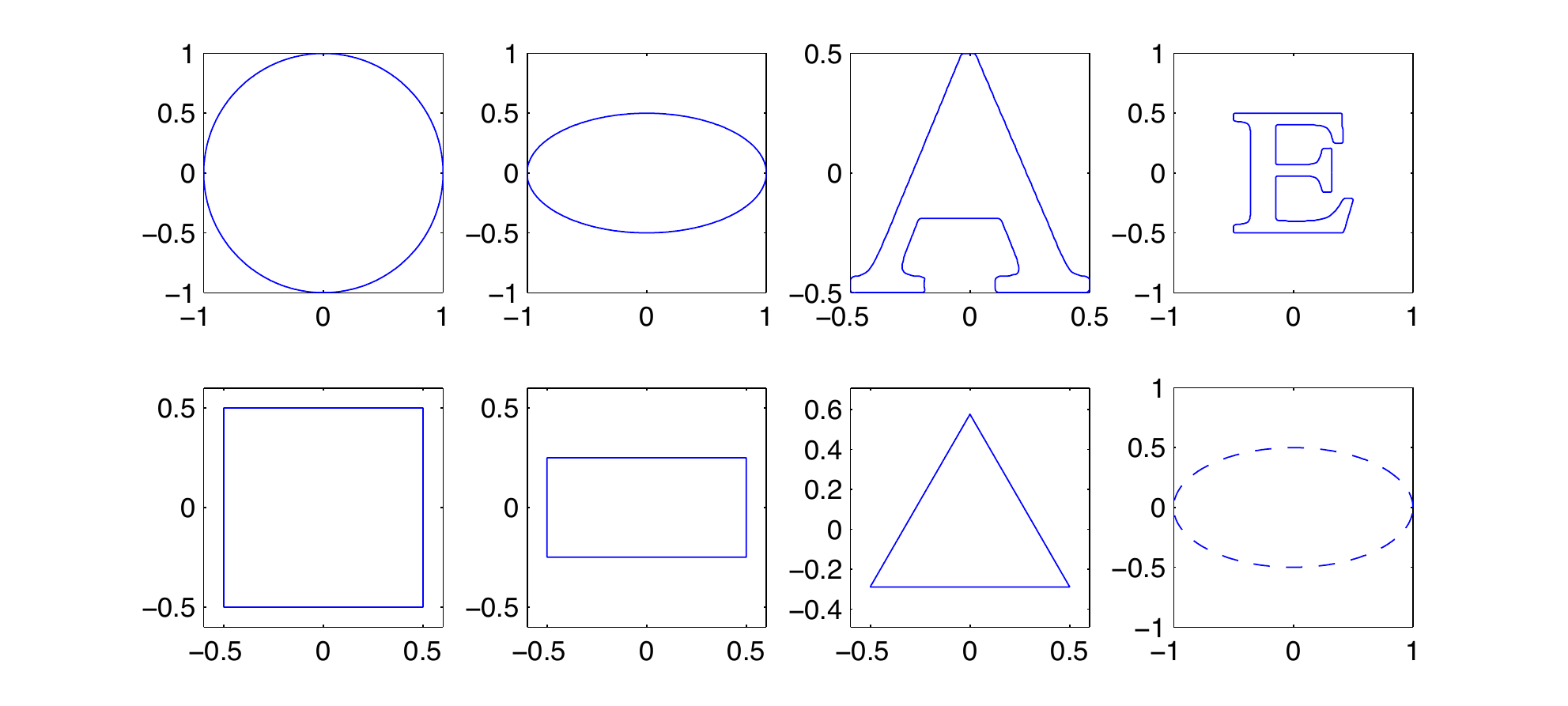}
	\caption{A dictionnary of shapes, used to identify the object $D$. The distance units here are arbitrary. Note that these shapes are often encountered in experimental studies~\cite{von2007distance} showing that the fish are able to recognize shapes.}
	\label{fig:dico}
\end{figure}

To solve this inverse problem we got inspired by the behavioral studies (for example~\cite{von1998electric}) described in the introduction. Indeed, instead of trying to \emph{compute} a shape from the measurements, we wanted to use classification and machine learning techniques. For example, let us suppose that the fish has already encountered several shapes $D_l$, $p=1,\ldots, L$ (such as the shapes plotted in Figure~\ref{fig:dico}), and that it knows all the polarization tensors ${\bf M}(k_{D_l}(\omega_j),D_l)$, where $\omega_j$ are the different frequencies emitted by the electric organ.

\begin{figure}[!ht]
	\centering
	\includegraphics[width=12cm]{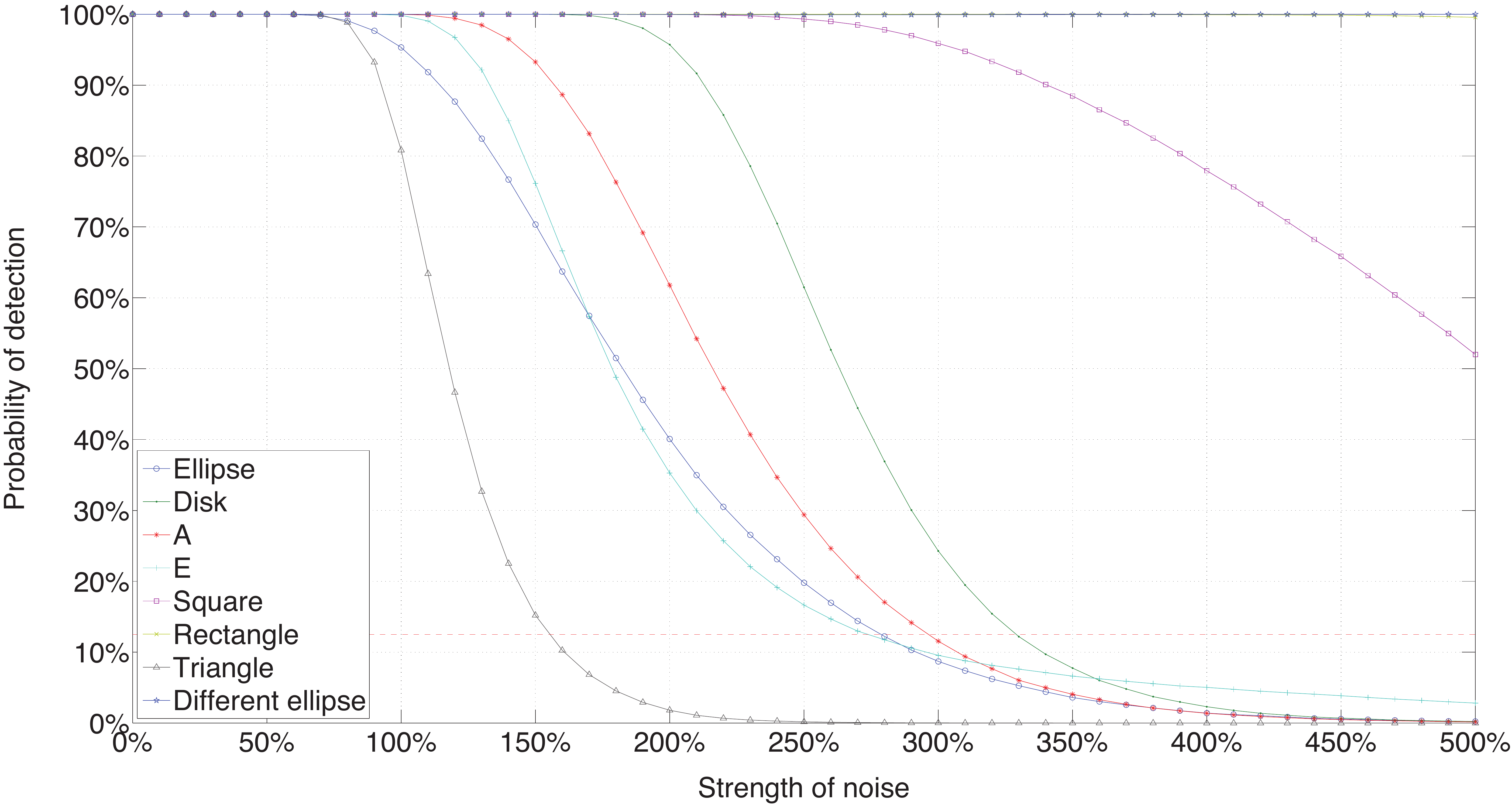}
	\caption{Performance of classification with respect to measurement noise. Each point represent the probability of correct detection, infered after $10^5$ realisations of the same experiment, consisting of the fish swimming around the object on a circular trajectory. The first-order polarization tensors $\MkD$ were extracted for $10$ frequencies (from $1\textrm{kHz}$ to $10\textrm{kHz}$). Then these tensors where compared to the dictionnary presented in Figure~\ref{fig:dico}. On the $x$-axis, the strength of the measurement noise is indicated, in percentage of $\norm{\ddn{u}-\ddn{U}}{}$. Measurement noise was modeled as a white Gaussian random process for each point, with standard deviation as what we call the \emph{strengh}. Originally published in~\cite{ammari2014shape}.}
	\label{fig:pnas}
\end{figure}

It is not possible to extract $\MkD$ from one single measure of $\ddn{u} - \ddn{U}$; instead, we use the fact that the fish actively swim around their prey when hunting, leading to particular swimming patterns called \emph{probing motor acts}~\cite{toerring1979motor}. Extracting ${\bf M}(k_D(\omega_j),D)$ from (\ref{eq:dipolar_formula}) measured for several fish's positions
is then a simple linear system~\cite{ammari2014shape}. Hence, $D$ can be identified as 
\begin{equation}
	D = \textrm{arg} \min_{D_l} \sum_j \norm{{\bf M}(k_D(\omega_j),D) - {\bf M}(k_{D_l}(\omega_j),D_l)}{}.
\end{equation}
In Figure~\ref{fig:pnas}, one can see the performance of this technique in terms of robustness against measurement noise.
Note that the number $64$ of electrodes is not very large in the numerical simulations,
but $10$ frequencies and $20$ different positions of the fish around the object are exploited:
the different positions allow for good extraction of the polarization tensors, and the different frequencies allow for good classification
of the object given the estimated polarization tensors. 
That is how robustness is achieved, and this remark may be of interest for the design of EIT devices.

\section{Conclusion and Perspectives}
\label{sec:conclusion_and_perspectives}
In this short review, we aimed at summarizing our main results on the mathematical modelling of active electrolocation. After deriving the partial differential equations and their boundary conditions, we have shown numerical simulations of the forward problem. Then, based on the dipolar approximation, we have detailed our localization algorithm and our shape recognition process. These latters are made possible thanks to multi-frequency measurements. Moreover, shape recognition additionally needs movement of the fish.

Note that our model is an extremely simplified version of what actually happens in real life; our aim was to extract the relevant features of the problem in order to be able to make a generalization for other topics, such as medical imaging or robotics. More realistic simulations and algorithms of target location active electro-sensing can be found for example in~\cite{babineau2006modeling,babineau2007spatial,lewis2001neuronal}. Shape identification remains an open challenge, although some algorithms begin to emerge in the robotic community~\cite{bai2015finding,lanneau2016object}.

Since the publication of our two algorithms, several directions of research have been taken. First, we have extended the shape recognition algorithm to the time-domain formulation of the problem~\cite{ammari2016time}, and to the echolocation problem (as observed in dolphins and bats for example)~\cite{ammari2014shapeecho}. Then, we have used wavelets methods in order to improve the accuracy of recognition~\cite{ammari2016wavelet}. And finally, we have raised the question of tracking a moving object~\cite{ammari2013tracking}.

\subsection*{Acknowledgement}
The authors would like to thank the reviewers for their very constructive remarks.

\bibliographystyle{plain}
\bibliography{biomimetics}

\end{document}